# Time domain deBroglie wave interferometry along a magnetic guide

Saijun Wu[1,2], Edward J. Su[1], Mara Prentiss[1]

1. Department of Physics and Center for Ultra Cold atoms, Harvard University, Cambridge, MA, 02138

2. Division of Engineering and Applied Science, Harvard University, Cambridge, MA, 02138

**Abstract:** Time domain deBroglie wave interferometry [Cahn et al, Phys. Rev. Lett. 79, 784] is applied to $Rb^{87}$ atoms in a magnetic guide. A standing wave light field is carefully aligned along the guiding direction of the magnetic trapping potential from a soft-ferromagnetic 4-foil structure. A sequence of two standing wave pulses is applied to the magnetically trapped atoms. The backscattered light at the atomic density grating revival time is collected and detected via a heterodyning technique. In addition to the observed recoil oscillations that fit the interferometer theory for atoms in free space, we observe a decay of the interferometer contrast on a millisecond time scale with unexpected millisecond-scale oscillations. We find that the oscillating decay is explained by a residual variation of the linear trapping potential along the standing wave direction.

## I. Introduction:

An ability to coherently control the motion of atoms close to a surface may eventually lead to the realization of chip-based atom interferometric devices for precise measurements and quantum computing, which are believed to be favorable over their free-space version due to their compactness and robustness. Although significant progress has been achieved in developing the building blocks of these devices, in most cases the external motion of the trapped atoms in the chip potential can be modeled using only classical mechanics[1-6]. A major difficulty associated with chip based atom interferometry is the realization of a chip-based beamsplitter that generates a mutual coherence between the trapped atoms. Various chip-based beamsplitter schemes have been proposed such as those based on tunneling [7], adiabatic followings [8, 9], and diabatic projections [10] of atomic waves. Generally speaking, a potential used for splitting the trapped atom wavepackets favors sharp features with sizes comparable to or smaller than the de Broglie wavelength of the trapped atoms, requiring either very cold atoms or potentials which vary substantially over a micron [11]. The size scale of the variation in the potential due to an atom chip can be reduced by trapping atoms closer to the chip, but bringing atoms closer to the chip surface can reduce interferometer performance because unwanted potential variations due to fabrication errors become more important and the coupling of the atoms to various surface noise channels increases [12, 13]. The problem is easily circumvented if the splitting potential is instead generated by an optical standing wave above the chip surface, as demonstrated in a recent experiment, where standing wave pulses coherently manipulated the guided motion of a Bose-Einstein condensate on a chip [14].

A Bose condensed atom sample provides a sample size limited coherence length that may be useful for interferometry; however, the atom-atom interactions of a condensate contribute additional challenges to an interferometric experiment. In this paper we describe experiments using a trapped atom sample that is far from quantum degeneracy, where we study the motion of the atoms along a magnetic guide using the time-domain de Broglie wave interferometry technique [15, 16]. At short times scales (<< 1ms), our interferometer signal shows recoil oscillations in the contrast of the $\lambda/2$ atom density grating due to the interference of guided atom wavepackets with themselves, which precisely fits the predictions of interferometer theory for atoms in free space. On the millisecond time scale we see an oscillatory decay of the interferometer contrast that is sensitive to the standing wave – magnetic guide direction misalignment angle, indicating its origin from the influence of the residual trapping potential along the standing wave direction. In this experiment, a fairly large spreading of our atomic sample in the slightly curved guiding potential has made a perfect match of the standing wave – guide direction across the atomic sample impossible. Nevertheless, by carefully adjusting the direction of the standing wave, we obtain a trap-strength dependent maximum interferometric interrogation time of 1-3 milliseconds for the magnetically confined atoms. We further suggest the improvement over this work with atomic sample localized along a straighter guiding potential. The goal of our future work along the path is to coherently manipulate the guided motion of atoms with light pulses over a time scale significantly longer than the oscillation period of the atoms in the guide. The technique would be important for the realization of chip-based atomic interferometric devices, and may also provide the opportunity to surpass the interrogation time limits for free-space atom interferometers, which, unlike their guided counterparts, are limited by the ballistic motion and free fall of atoms.

The remainder of the paper is organized in 3 parts. In part II we briefly review the theory of a time domain de Broglie interferometer, and discuss the generalization of the formula to a magnetically trapped atom sample. A formula that describes the leading order influence of trapping potentials on the interferometer contrast decay is then derived. In part III, we describe the experimental setup, and in part IV we present the experimental results.

## II. Time domain de Broglie wave interferometer with trapped atoms

In this section we will briefly review the principle of the grating echo technique that leads to the time domain de Broglie wave interferometer as was first demonstrated by Cahn et al [15], we will then discuss the application of the free-space interferometer theories to atoms confined in a magnetic guide.

An off-resonant standing wave light pulse induces two-photon transitions for a ground state atom, the result is to put the atom into a coherent superposition of different diffraction orders with momentums shifted by integer multiples of $\hbar\Delta\mathbf{k}$, where $\Delta\mathbf{k}$ is the wave vector of the standing wave. Shortly after the standing wave potential is applied, the different diffraction orders interfere constructively to generate density grating with spatial frequencies that are integer multiples of $\Delta\mathbf{k}$ (atomic Talbot effect). The contrast of the density grating falls off rapidly as the density gratings formed by atoms with different initial velocities becomes more and more displaced in space. However, a revival of the atomic density grating is achievable at the so-called population grating echo time after the standing wave is pulsed twice [15, 16]. In the work described by [15] this light pulse echo technique was first implemented for an interferometric measurement, and is referred as a time domain de Broglie wave interferometer.

In the time domain de Broglie wave interferometer, a cold atom sample is subjected to two standing wave light pulses with pulse area (Integrated atomic ground state light shift over pulse duration) $\theta_1$ and $\theta_2$ separated by time T, before a probe light with the wave vector $\mathbf{k_1}$ is switched on at time 2T+t to probe the atomic density grating by monitoring the back scattered light with the wave vector $\mathbf{k_2}$. The backscattered light field $E_s$(2T+t) is shown to be proportional to the magnitude of the atomic density grating measured by $\rho_{\Delta k}$, the $\mathbf{\Delta k = k_2 - k_1}$ Fourier component of the atom density distribution at the probe time 2T+t, e.g.,

$$E_s(2T+t) \sim \rho_{\Delta \mathbf{k}}(T,t) \tag{1}$$

$\rho_{\Delta \mathbf{k}}(T,t)$ can be derived by integrating the equation of motion for single atoms subjected to the standing wave pulses at time 0 and T followed by two free evolution period from 0 to T and from T to 2T+t, Before the assemble average is taken out on the spatial probability distribution for each atom trajectories. For short light pulses in the Raman-Nath region such that the atomic motion is ignorable during the pulse duration, and for a initial atom sample with spatial spreading much larger than the light wavelength and a thermal deBroglie wavelength much smaller than the light wavelength, $\rho_{\Delta k}$ is given by:

$$\rho_{\Delta \mathbf{k}}(T,t) \approx \rho\, \theta_1 J_2[2\theta_2 Sin(4\omega_r T)] t e^{-(\Delta k u t/2)^2} \tag{2}$$

Here $\rho$ is the density of the atom sample projected along the $\mathbf{\Delta k}$ direction, $u$ is the mean thermal velocity of the atoms, $\omega_r$ is the angular recoil frequency of the atom defined as $\omega_r = \frac{1}{4}\frac{\hbar \Delta k^2}{2m}$ where m is the mass of the atom. The Fourier component $\rho_{\Delta \mathbf{k}}(T,t)$ depends on t<<T through the factor $t e^{-(\Delta k u t/2)^2}$ as a dispersion like peak close to the echo time t=0, with its width proportional to the thermal deBroglie wavelength of the atomic sample, and amplitude dependent on the interrogation time 2T through the second order Bessel function $J_2[2\theta_2 Sin(4\omega_r T)]$.

The expression of the backscattered light field given by (1) and (2) is derived by assuming the atomic motion is only subjected to the pulsed light shift potentials and would otherwise be free. Our interferometer experiment is conducted with atoms confined in a magnetic guide near the field strength minimum of a magnetic 2D quadruple field (See fig 1). The standing wave direction, e.g., the direction of $\mathbf{\Delta k}$, is carefully aligned to be parallel with the magnetic guiding direction. Thus ideally we expect (2) to be valid for our experiments since the magnetic confinement is orthogonal to the motion of atoms in the standing wave direction along which the interferometer experiment is taken out. Practically, however, the confining potential variation along $\mathbf{\Delta k}$ direction cannot be completely eliminated, not only due to the non-ideal standing wave alignment, but also due to the fact that the guiding potential itself is neither perfectly straight nor translational invariant. Thus it is helpful to derive an expression of $\rho_{\Delta k}(T,t)$ that takes into account the perturbation from the residual confining potential variations.

To derive the expression, let's first consider the simple case where the motion of atoms is perturbed by a uniform linear potential $V(\mathbf{r}) = -m\mathbf{a}\cdot\mathbf{r}$. It can be shown that without altering the contrast, the presence of the uniform acceleration displaces the final atom density grating by $\delta\mathbf{r} = \frac{1}{2}\mathbf{a}(2T)^2$. Thus (2) is modified by including a phase factor $e^{i\mathbf{\Delta k}\cdot 2\mathbf{a}T^2}$, e.g.,

$$\rho_{\Delta k}(T,t) = \rho\,\theta_1 t J_2[2\theta_2 Sin(4\omega_r T)] e^{-(\Delta k u t/2)^2} e^{i2\mathbf{\Delta k}\cdot\mathbf{a}T^2} \qquad (2')$$

The influence to the atomic density grating from a general potential V(**r**) cannot be formulated as simply as in (2'). If, however, V(**r**) is smooth enough such that the local acceleration $\mathbf{a}(\mathbf{r}) \equiv \frac{\nabla V(\mathbf{r})}{m}$ is approximately a constant along the standing wave direction $\Delta \mathbf{k}$ for most of atoms during the interrogation time 2T, we may ignore the curvature of the potential for any individual atom trajectory, and retrieve $\rho_{\Delta k}(T,t)$ in a semi-classical picture by spatially averaging the final probability distribution of atoms at different locations according to (2') with $\mathbf{a} = \mathbf{a}(\mathbf{r})$. Thus we end up with,

$$\rho_{\Delta k}(T,t) = \rho\, \theta_1 t J_2[2\theta_2 Sin(4\omega_r T)] e^{-(\Delta k u t/2)^2} < e^{i 2 \Delta k \bullet \frac{\nabla V}{m} T^2} >_{\mathbf{r}} \tag{3}$$

To apply (3) to the experiment that will be discussed in this work, consider an atom with magnetic moment $\mu$ confined in a 2D magnetic guide with magnetic field in Cartesian coordinates that is given by $\mathbf{B} = (B_1 y, B_1 x, B_0)$. The confining potential can be written as V(**r**) $= ma\sqrt{x^2 + y^2 + r_0^2}$ with $ma = \mu B_1$ and $r_0 = \frac{B_0}{B_1}$. Consider a standing wave with a wave vector $\Delta \mathbf{k} = \Delta k(\delta, 0, 1)$, where $\delta \ll 1$ is the misaligned angle between the guiding direction $\mathbf{e}_z$ and the standing wave direction $\Delta \mathbf{k}$. Assume the trapped atom sample has a Gaussian distribution with a width $\sigma$ close to the magnetic field minimum, we have:

$$< e^{i 2 \Delta k \bullet \nabla V T^2} >_{\mathbf{r}} = \iint dx dy \frac{1}{\pi \sigma^2} e^{-\frac{x^2 + y^2}{\sigma^2}} e^{i 2 \delta \Delta k a T^2 \frac{x}{\sqrt{x^2 + y^2 + r_0^2}}} \tag{4}$$

Integrate (4) in cylindrical coordinate gives:

$$< e^{i 2 \Delta k \bullet \nabla V T^2} >_{\mathbf{r}} = \int_0^\infty e^{-u} J_0(2\delta \Delta k a T^2 \frac{1}{\sqrt{1 + \frac{r_0^2}{\sigma^2 u}}}) du \tag{4'}$$

Where $J_0$ is the 0$^{th}$ order Bessel function.

The integration of (4') becomes particularly simple for a very small or very large ratio $\frac{r_0}{\sigma}$. If most atoms are in the harmonic region near the bottom of the guide (harmonic trap limit), e.g., $\sigma \ll r_0$, (4')

can be approximated with

$$< e^{i2\Delta k \bullet \nabla V T^2} >_r \approx e^{-(\frac{\delta\sigma\Delta ka T^2}{r_0})^2} \tag{5}$$

If instead most of the atoms are in the linear region of the guide (linear trap limit), e.g., $\sigma \gg r_0$, we end up with:

$$< e^{i2\Delta k \bullet \nabla V T^2} >_r \approx J_0(\frac{\delta\Delta ka(2T)^2}{2}) \tag{5'}$$

Combined with (3), we see that the misalignment of the standing wave with respect to the guiding direction results in the decay of the interferometer contrast with respective to the interrogation time 2T, with the contrast decay factor given by (5) or (5') in either limit. In the harmonic trap limit, the decay actually has a characteristic time scale of $\sqrt{\frac{1}{\delta\Delta k\sigma} \frac{2}{\omega_0}}$, where $\omega_0$ is the transverse angular trapping frequency of the harmonic guide. The experimental situation in this work turns out to be closer to the linear trap limit, where (5') predicts an oscillatory decay that is intuitively understood as the interference of contributions to $\rho_{\Delta k}(T,t)$ from atoms at different sides of the linear trap. The decay can be characterized by $2T_0 \approx \frac{2.2}{\sqrt{\delta\Delta ka}}$, with which the argument in (5') equals the first zero of the 0$^{th}$ order Bessel function and the contrast modulation meets its first node. Although both (5) and (5') break down for large 2T that becomes comparable with the transverse atom oscillation period (since the residual potential curvature cannot be ignored for single atom trajectories anymore), the minimization of the misalignment could in principle lead to an interferometer interrogation time 2T longer than the transverse oscillation period of trapped atoms in the guide and thus realizing a guided atom interferometer. Of course, such an elimination of the misalignment requires the guiding potential itself to be straight and homogeneous across the atomic sample along the guiding direction.

Two notes may be added to the above treatment of interferometric contrast decay given by expression (5) or (5'): First, these expressions give a contrast decay factor due to inhomogeneous phase shifts of the atomic ensemble in the external field. The expression is derived for the time-domain de Broglie interferometer here, but is applicable to general interferometer configurations. Second, we have assumed the atomic Zeeman shift and the light shift can simply be added up together. This is not generally true due to the tensorial nature of the light shift potential [17]. The nonlinear magneto-optical effects could introduce extra complications in our interferometer experiment such as an inhomogeneous pulse area across the atom sample and a loss of trapped atom sample due to spin flips. These effects influence our experimental results negligibly. We also notice that these effects can in principle be eliminated for alkali atoms by choosing linearly polarized far-off - resonant light [17].

## III. The Experimental setup

### III.1 In situ loading of atoms in a magnetic guide[18]

We use a 4-foil magnetic structure to generate the 2D quadruple magnetic field for the $2D^+$ magneto-optical trap as well as the magnetic confining potential. The structure is composed of four 0.5mm thick, 65mm long by 31mm wide rectangular $\mu$-metal foils that are placed parallel to each other with 5mm separations (see fig.1a). The current sheets that run through the wires around the foils pull the magnetization of the 4 foils up and down alternatively to generate a 2D quadruple magnetic field on top of the 4-foil structure. By increasing the magnetization of the two inner foils and the two outer foils in proportion, the magnetic field gradient at the 2D field strength minimum can be varied while the position of the minimum remains fixed.

About $10^8$ $Rb^{87}$ atoms are cooled and trapped from the background vapor by a $2D^+$ magneto-optical trap (MOT), 6mm away from the foil structure. In the last 12ms of the MOT operation, first the magnetic field gradient ramps from 20G/cm to $B_1$ (up to 100G/cm) in 10ms, while the cooling laser intensity is reduced and detuned for polarization gradient cooling. The repumping laser is switched off in the last 1ms before the MOT light is switched off, resulting in approximately 3 $\times$ $10^7$ atoms in F=1 hyperfine states trapped in the magnetic potential. The transverse width of the atom distribution is around 200 $\mu$ m at a magnetic gradient $B_1$=50G/cm, and is inversely proportional to $B_1$, indicating that most of atoms has a thermal distribution in the linear trapping region of the confining potential [19]. The trapped atoms have a mean longitudinal velocity u ~ 5cm/s that is roughly 8 times the recoil velocity of the atoms. After the atoms have been transferred to the magnetic trap, we wait for a time $t_{esc}$ for the untrapped atoms to escape the trap region, and perform the standing wave experiment with the magnetically trapped atoms. $t_{esc}$ has been sampled from 3ms to 30ms, where $t_{esc}$>9ms is proved to be enough to eliminate the contribution of residual untrapped atoms to the back scattering signal. As shown in fig 1.b, the motion of atoms confined in the 2D magnetic trap can be studied by taking absorption images from the side of the trap in repeated experiments. (The untrapped atoms are with a smaller optical depth due to large spreading, and are not visible in the images shown in fig1.b)

**III.2 The standing wave experiment**

The standing wave is composed of two traveling laser beams that overlap at the atom trap region with an intersection angle $\Theta$ around 90mrad. The two beams, which will be referred as $I_1$ and $I_2$, are independently controlled by acoustic-optical modulators, and have a $1/e^2$ diameter of 4mm and 2mm respectively. The standing wave light is far blue detuned relative to the $Rb^{87}$ D2 F=1 – F'=2 transition (+136Mhz in this experiment). While keeping the direction of $I_2$ fixed, the direction of $I_1$ can be adjusted to tune the standing wave direction relative to the magnetic guide direction (see fig.2).

Two 600ns standing wave pulses separated by time T are applied to the atom sample. The pulse area for both pulses is estimated to be $\theta$=2.5. $I_1$ is then switched on as the probe light from 2T-4$\mu$s for more than 100$\mu$s, during which the backscattered light is continuously collected with the detecting optics. In a time-domain deBroglie wave interferometer experiment, the backscattered signal light shares the same propagation direction as one of the traveling wave beams forming the standing wave. The detecting optics setup requires extra care due to the fact that the traveling wave beam can be orders of magnitude stronger than the signal light. In our experiment (see fig.2), a 40Mhz AOM is used as a protective shutter before the detecting photodiode, and is switched on only when $I_2$ is switched off. Further, to detect the nanowatt level signal light, $I_2$ is attenuated more than 70dB by switching off its controlling AOMs during the probe time.

The backscattered light is mixed with an optical local field at a different frequency and collected with a single-mode optical fiber that delivers the light to a fast avalanche photodiode (APD). The mixed electronic signal from the APD is further mixed down to 40Mhz, and put into a Tektronix digital oscilloscope. The amplitude of the backscattering signal is retrieved by using the oscilloscope's native FFT to measure the component of the electronic signal at frequencies near 40Mhz. The standing wave experiment setup is summarized in fig.2; we also show a typical FFT signal from a single iteration of the experiment. The retrieved FFT double-peak in fig.2 corresponds to the modulus of the Fourier transformed time-domain signal described by (2). The time domain deBroglie interferometer is completed by varying the interrogation time 2T in repeated experiment while recording the peak value of the FFT signal.

## IV. Results and discussion

We begin the description of our experimental results with a data set that samples 2T from 0.2ms to 0.6 ms in 2 $\mu$ s separations (Fig. 3a). The standing wave experiments are taken out with $B_1$=20G/cm and $t_{esc}$ =3ms. Since the magnetic field is unaltered during the whole experiment, the experiment can be repeated with a repetition rate as high as 5Hz. The recoil oscillation in Fig. 3 a) agrees well with the theory described in part II, e.g., $<|E(2T, t)|>_t \sim |J_2[2\theta Sin(4\omega_r T)]|$. The optical pumping effect can be included in the expression (2) in part II by including a small imaginary part to the pulse area $\theta$ [20]. Fig 3d) shows a simulation result on the experimental curve based on the simply model with $\theta$ set to be 2.6 + 0.08i. A comparison between the experimental curve in fig 3.b) and the simulation curve in fig 3.d) shows that on a short time scale the backscattered signal is well described by the free-space interferometer model. A linear fit of the minimum backscattering signal time $T_n$ with the recoil phase n can be used to retrieve $\omega_r = \frac{1+\cos\Theta}{2}\omega_r^0$, where $\omega_r^0 = 2\pi \times 3.771$ kHz is the D2 line recoil frequency of $Rb^{87}$ atom. The intersection angle $\Theta$ between $I_1$ and $I_2$ has been measured to be 80 mrad in this experiment. The fit in fig.3 b) yields the measured recoil frequency $\omega_r^0|_{\exp} = 2\pi \times (3.771 \pm 0.001)$ kHz. The result is consistent in different measurements with different intersection angles $\Theta$.

Given the time dependence of the backscattered signal derived above and the precise measurement of the period of the oscillation, we can ignore the sub-millisecond scale recoil oscillations and directly retrieve the longer time scale behavior by sampling the peak contrast point for each oscillation. In particular, we see from fig3.a that at $2t_n = 2T_n + 8\mu$s the backscattering signal reaches a peak in each recoil oscillation period. By sampling the backscattering signal with interrogation time $2T$ around $2t_n$, we were able to study the millisecond scale behavior of the interferometer contrast with a small number of sampling points. The millisecond-scale experiments were done with atoms prepared in different trapping potentials $B_1$. In addition, the standing wave direction was tuned relative to the magnetic guide direction by adjusting the $I_1$ direction both horizontal and vertically. Typical experimental results are shown in fig. 4 and 5. The oscillatory decay of the backscattering signal is clearly seen in all the data. With the formula expressed in (5'), we characterize the oscillatory decay of the backscattering signal in fig. 5 with $2T_0$, the interrogation time with which the backscattering signal amplitude meets its minimum. In fig.6 a, we show that $B_1 T_0^2$ is approximately a constant as predicted by expression (5') through the relation $2T_0 \approx \dfrac{2.2}{\sqrt{\delta \Delta ka}}$. We also calculate the quantity $\delta$ (which, with the model described by (5'), should measure the misalignment angle between the standing wave direction and the guide direction) through the relation for each set of the millisecond scale data. Each recorded data set can be represented by a point on the ($\delta, \Theta$) plane, as shown in fig. 6b that includes the recorded data with $I_1$ scans approximately horizontal along the $I_2$

– guide plane. Notice the variation of $\delta$ scales by roughly half that of the variation on the intersection angle $\Theta$, as expected from the relation **$\Delta$k =k$_2$-k$_1$**. However, we have not yet been able to reduce $\delta$ to less than 5mrad. This is unsurprising – the magnetic guide generated by the 4-foil structure is of course not perfectly straight. The edge fields from the ends of the 4-foils effectively bend the magnetic guide vertically so that the guiding direction varies along a trapped atom sample. The curvature of the guide in this experiment has been shown to be ~1m$^{-1}$ in the magnetic field simulations. Thus the guiding direction should vary up to 5mrad for a 5mm atom sample along the guide, which could explain our inability to completely eliminate the effects of misalignment.

To confirm that the oscillation feature in the backscattered signal is due to the magnetically trapped atoms, we conducted a comparison experiment, where the trapped atoms were moved to 1mm below the standing wave region by ramping up the magnetic bias field during the escape time. As shown in fig. 7, the resulting amplitude of the backscattering signal is below the noise level, indicating after an escape time of 10ms the residual untrapped atoms make a negligible contribution to the trapped atom interferometer experiment. In fig 7 we also included an interferometer contrast signal conducted with atoms prepared in free-space (with a variation of the magnetic field gradient at the standing wave region estimated to be less than 1G/cm.). Compare with the trapped atom case, we see that the contrast decay with respective to 2T: a) is monoclinic and actually matches the time scale of the free-space atomic density decay in the standing wave interrogation region, and b) is obviously slower than that have been achieved in all the experiments taken out with trapped atoms so far in this work. The fast decay of interferometer contrast in the trapped atom case should be able to be suppressed with a better match of standing wave direction with the field strength invariant direction (guide direction) at the interrogation region. This requires a sub-millirad guiding direction variation across the atomic sample, which should be achievable with an atomic sample in a straighter guide with a smaller spreading along the guiding direction (not necessarily to be transversely localized).

It is also worthwhile to compare our results with the recent guided atom interferometer experiment conducted with a BEC [14]. First, we notice that the inhomogeneous interferometric phase shift due the wide spreading of the atomic sample initial phase space distribution has not prevented us from conducting an interferometer experiment with atoms in a magnetic trap, with the achieved interrogation time comparable with the work in [14]. This is due to the implementation of the population grating echo technique [15] that has made the interferometric phase shifts here less sensitive to the initial velocities of atoms.

Secondly, the contrast decay factor derived in part II of this paper may also be applied to estimate the contrast decay in [14], where the guiding potential obviously reaches the harmonic trap limit so that the factor (5) has to be applied instead of (5'). In contrast to this experiment, the atomic sample and the guiding potential in [14] is ideal for a precise standing wave alignment since the atomic sample has a very small initial spreading along the guiding potential with a negligible curvature. Unfortunately, both the direction of the guide and the standing wave direction in [14] have been fixed during the fabrication of the atomic chip, with an uncertainty on the order of 2 degrees (30mrad). Estimation based on (5') gives a contrast decay time in [14] no longer than the transverse oscillation period (~5 ms) of the magnetic guide if the misalignment angle is more than 10mrad. Thus we suggest the "misalignment problem" may significantly contribute to the observed contrast decay in [14] as well. The significance of the misalignment may require the development of fabrication technique that precisely pattern the direction of the relative conducting wires to be orthogonal to the surface of the mirrors mounted on top.

Thirdly, as suggested by [21], the inter-diffraction-order atomic interaction is not ignorable, and may also contribute to the observed interferometer contrast decay. In contrast, this particular decoherence channel is negligible in our sample since it is about 3 orders of magnitude lower in atomic densities. In a simple multi-mode guide picture, the strength of mean-field interaction is reduced by a factor of order $<N>$, if $<N>$ modes are populated instead of single mode guiding. The attempt to minimize the influence of the mean-field effect, in combination with the requirement of a precise standing wave alignment across a finite sized atomic sample, suggests the multi-mode guiding of a longitudinally localized atomic sample to be favorable for the standing wave light pulse manipulation on the guided motion of atoms.

**V. Conclusion**

In conclusion, we have studied the time domain de Broglie wave interferometer with atoms in a magnetic guide. We show that it is possible, using atoms in a magnetic guide far from quantum degeneracy, to achieve interferometric coherences time on the order of those achievable in free space. We also demonstrate, theoretically and experimentally, that the interferometric contrast is extremely sensitive to any residual variation of the confining magnetic field along the standing wave direction. This implies that any light pulse guided atom interferometer designed to surpass free-space versions in coherence time must include a very straight confining potential, and a method for precisely aligning the standing wave along its guiding direction.

**Acknowledgement:**

This work is supported by MURI and DARPA from DOD, NSF, ONR and U.S. Department of the Army, Agreement Number DAAD19-03-1-0106.

generated in a cloud of cold atoms, PHYSICAL REVIEW A **66**, 023601, 2002

[17] Mathur BS and Happer W, Light Shifts in the Alkali Atoms, Physics Review, 171(1) 11, 1969

[18] Vengalattore M, Rooijakkers W, Prentiss M, Ferromagnetic atom guide with in situ loading, PHYSICAL REVIEW A 66 (5) 053403, 2002

[19] The local field effect by the $\mu$-metal foils has prevented us from freely introducing a homogenous plug field $B_0$ along the guiding potential. An experimental control of $B_0$ and thus the ratio $\frac{r_0}{\sigma}$ to obtain (5) and (5') could have been helpful to gain a clearer insight into the influence of the magnetic potential to the interferometer experiment discussed in the work.

[20] It can be shown that, by including an imaginary part to the pulse area $\theta$, the Bessel functions involve in the equation (2) follows following replacing rule: $J_n[2\theta Sin(4\omega_r T)]$ is replaced by $J_n[\sqrt{x^2-y^2}](\frac{x+y}{x-y})^{n/2}$, with x=2 Real $[\theta] Sin[4\omega_r T]$, y= 2 Imag $[\theta]\cos[4\omega_r T]$

[21] Olshanii M and Dunjko V, Interferometry in dense nonlinear media and interaction-induced loss of contrast in micro-fabricated atom interferometers, cond-matt/0505358

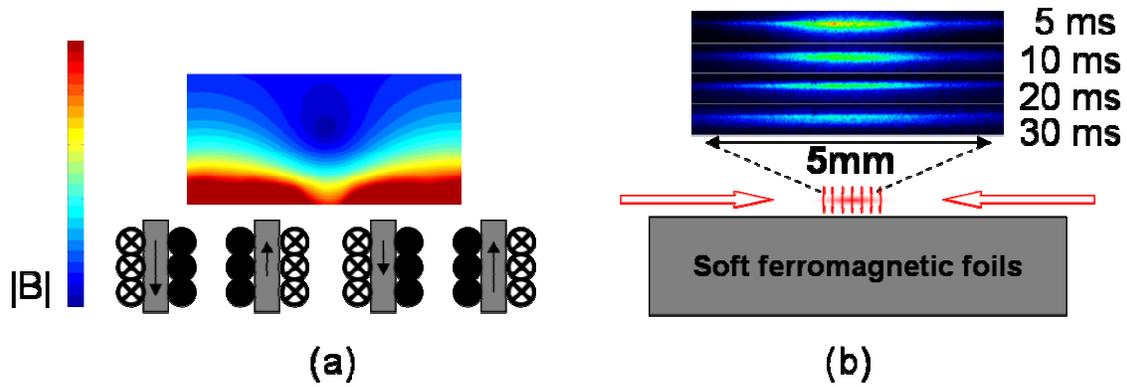

Fig.1 the 4-foil magnetic structure and the standing wave setup (scale not in proportion). (a) Cross-section of the 4-foil structure perpendicular to the guiding direction: a 2D quadruple field is generated by the four-foil structure, with the magnetization direction indicated by the black arrows inside each foil cross sections. The direction of the currents around each foil is also indicated. A calculated contour plot of the magnetic field distribution is shown on top of the structure. (b). Cross-section of the 4-foil structure along the guiding direction. On top of the structure the location of the atoms and standing wave are indicated. Four absorption images of the trapped atom sample taken at 5ms, 10ms, 20ms, and 30ms after the MOT light is switched off are shown on the top.

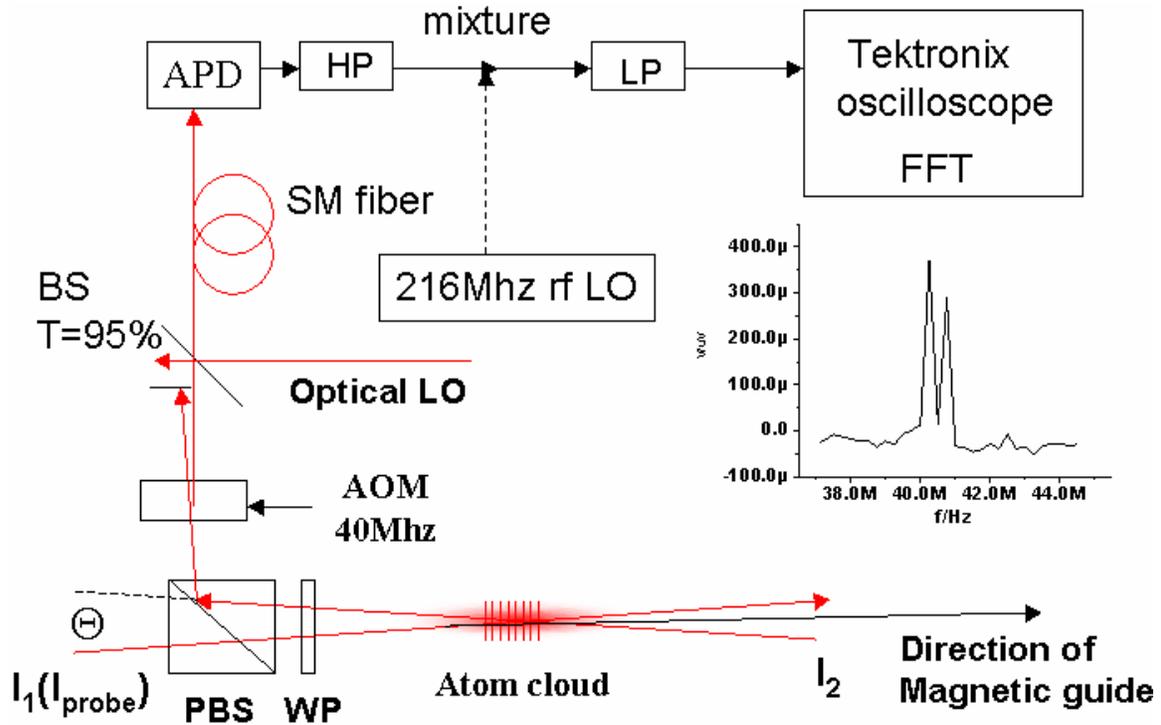

Fig.2 The interferometer setup. $I_1$ and $I_2$ intersect at the atom trap region to form a standing wave. The angle $\Theta$ between the two laser beams is around 90mrad, and is measured to 1% precision for each experiment. By adjusting the direction of $I_1$, the direction of the standing wave relative to the magnetic guiding direction can be adjusted. $I_1$ and $I_2$ are independently controlled via Acoustic-Optical modulators that are not shown here. An inserted graph on the middle-right is a typical fft signal from a standing wave experiment: WP: quarter-wave plate. PBS: polarization-dependent beamsplitter. BS: 95% transmission beam pickoff. SM Fiber: Single Mode fiber. APD: avalanche photodiode. LO: local field. HP: high band-pass filter. LP: low band-pass filter.

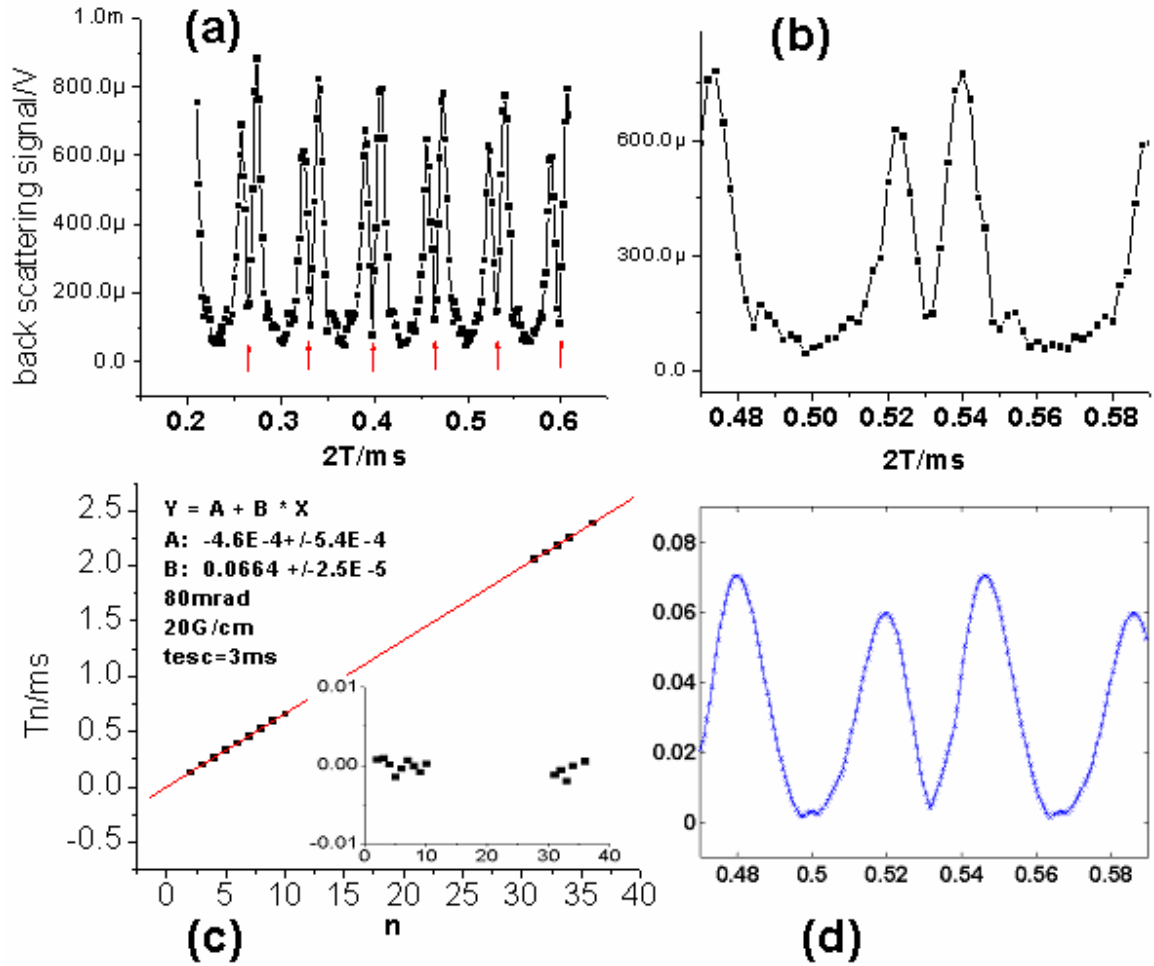

Fig. 3 The interferometer data taken with atoms in $B_1$=20G/cm 2D magnetic trap. The two traveling laser beams that make the standing wave intersect with an angle measured to be 87mrad. (a) The recoil oscillation, red arrows mark the time $T_n$ for which the backscattering signal reaches a minimum. (b) Expansion of the interferometer data around 0.52ms. (c) Fitting of $T_n$ with n, Inset gives the fitting error, which is smaller than 2 $\mu$s for all the measured $T_n$. (d) A calculated backscattering signal based on the model discussed in part II of main text.

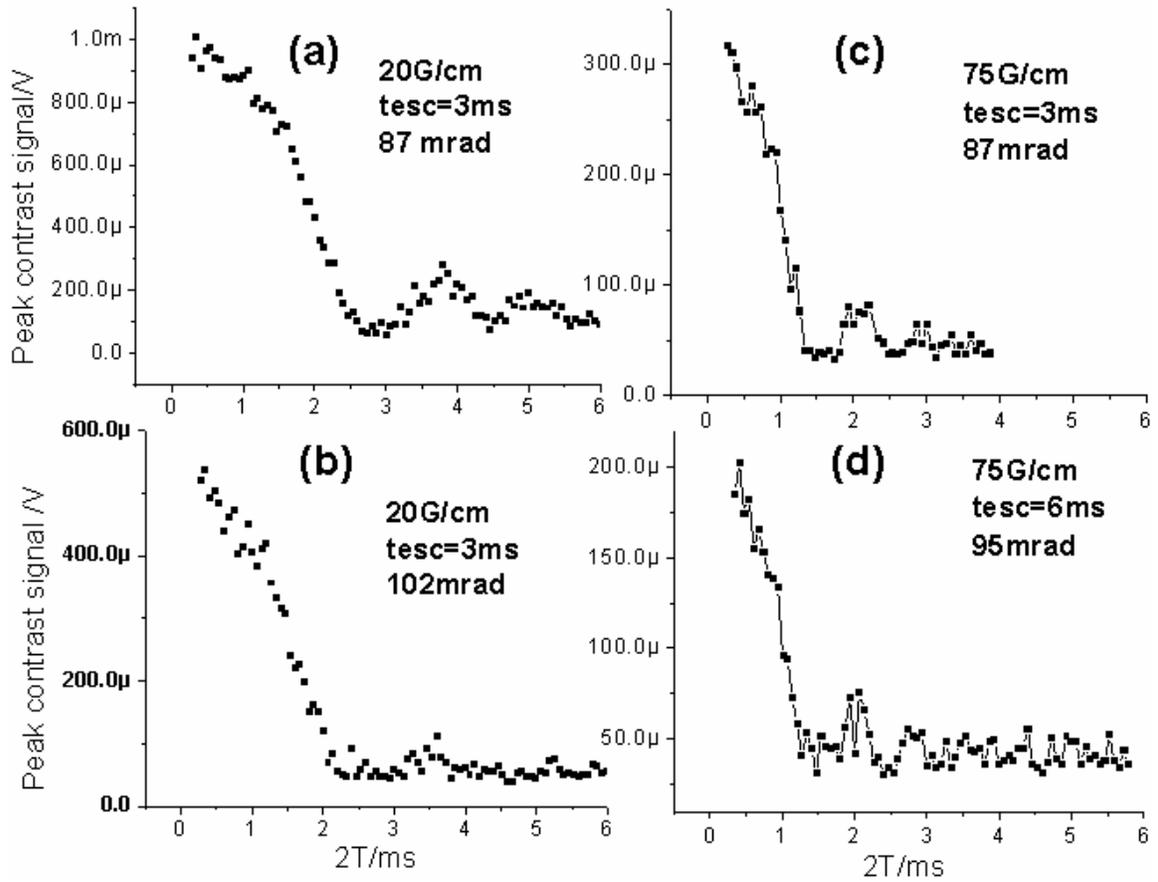

Fig.4. Millisecond-scale behavior of the interferometer signal (contrast decay). 2T is sampled at the peaks of each recoil oscillation period. (a) and (c) are taken with two laser beams intersect with a angle $\Theta$ = 87mrad. While in (b) $\Theta$ = 102mrad. (d) $\Theta$ = 95mrad. Different magnetic field gradient $B_1$ and the untrapped atom escaping time $t_{esc}$ are also indicated in each figure.

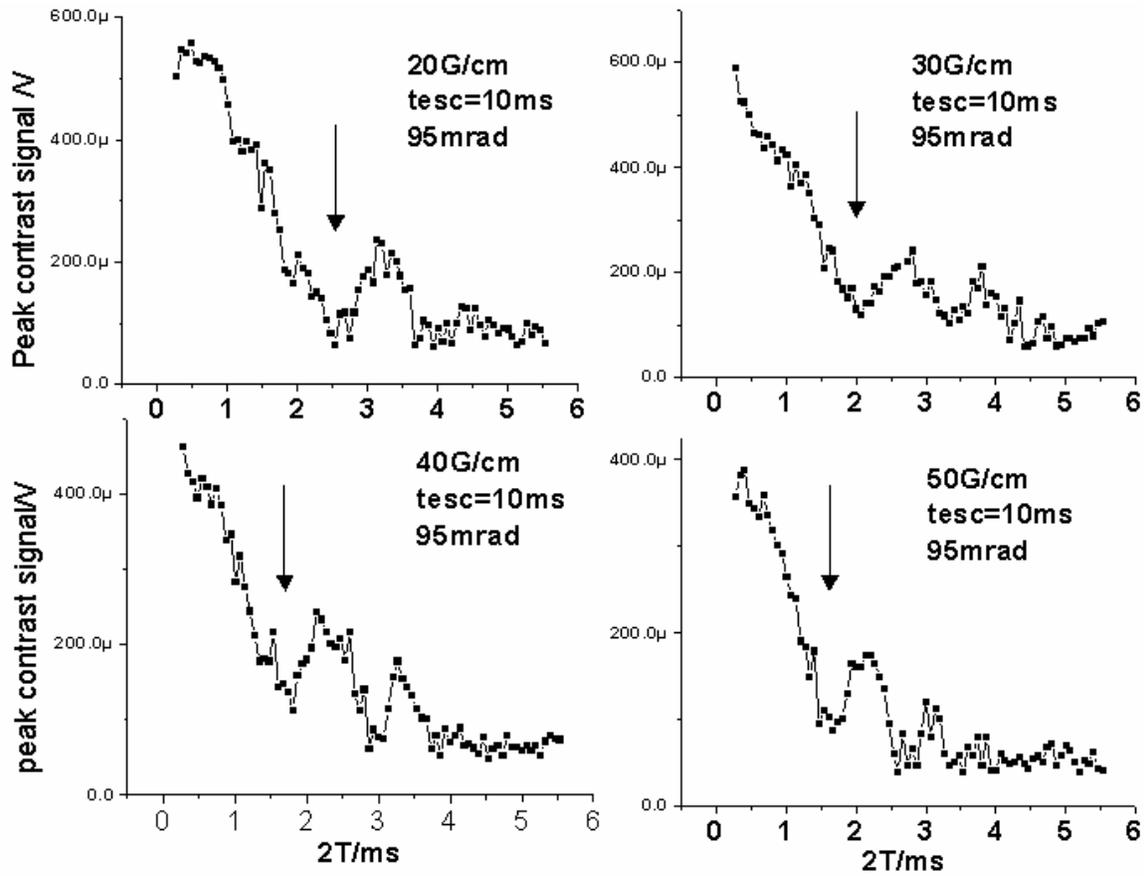

Fig. 5, The contrast decay in magnetic guide with different magnetic confinements. $t_{esc}$=10ms. $\Theta$=95 mrad. The arrow in each graph indicates the time $2T_0$ when the interferometer signal meets the first millisecond scale contrast minimum.

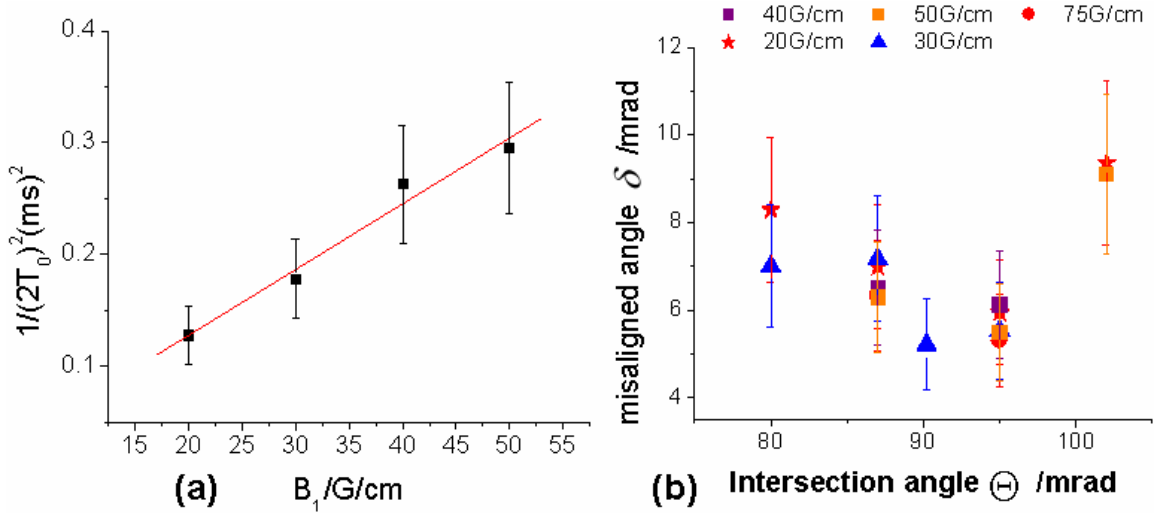

Fig. 6 (a) the linear relation between $1/(2T_0)^2$ and $B_1$ retrieved from data in fig. 5. The 10% uncertainty in the determination of $T_0$ is reflected in the error bars along the y-axis. Relative magnitude of $B_1$ at each data point is determined by the current run through the conductors, and have error bar smaller than the size of the symbols. (b) The misaligning angle $\delta$ is calculated from 14 set of data taken with $\Theta$ close to 90mrad, and plotted on the ($\delta$, $\Theta$) plane. Different symbols represent different $B_1$ values. The 10% uncertainty in determination of $T_0$ is reflected in the error bars of $\delta$. Intersection angle $\Theta$ is measured to 1% precision in each experiment and the errors are smaller than the size of the symbols here.

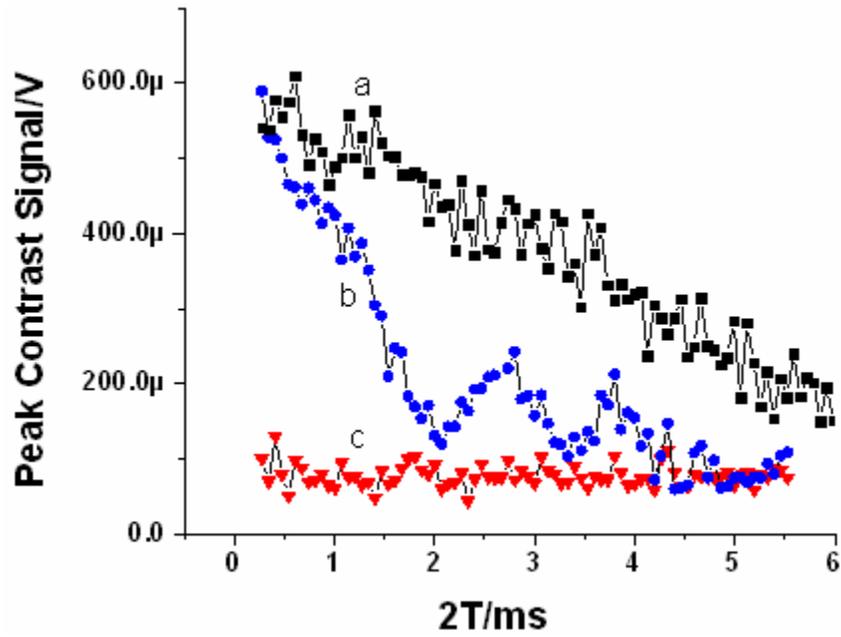

Fig. 7, Comparison of the interferometer contrast decay with atoms in free space (a, black dots); in presence of guiding potential $B_1$=30G/cm, $t_{esc}$=10ms. $\Theta$ =95 mrad and with (b, blue dot) and without (c, red triangle) the magnetically confined atoms. An averaging over 36 iterations has been applied to generate both curve b) and c), while in a) the average has only been taken over 10 iterations that explains the difference in signal/noise ratio.